\documentclass[journal]{IEEEtran}

\usepackage{glossaries}
\makeglossaries
\newacronym{gcd}{GCD}{Greatest Common Divisor}
\newacronym{lcm}{LCM}{Least Common Multiple}
\newacronym{ble}{BLE}{Bluetooth Low Energy}
\newacronym{iot}{IoT}{Internet of Things}
\newacronym{css}{CSS}{Chirp Spread Spectrum}
\newacronym{fsk}{FSK}{Frequency Shift Keying}
\newacronym{lpwan}{LPWAN}{Low Power Wide Area Network}
\newacronym{sf}{SF}{Spreading Factor}
\newacronym{crc}{CRC}{Cyclic Redundancy Check}
\newacronym{phy}{PHY}{Physical Layer}
\newacronym{mac}{MAC}{Media Access Control}
\newacronym{db}{dB}{Decibel}
\newacronym{bps}{bps}{Bits per second}
\newacronym{snr}{SNR}{Signal-to-Noise Ratio}
\newacronym{rf}{RF}{Radio Frequency}
\newacronym{ip}{IP}{Internet Protocol}
\newacronym{ism}{ISM}{Industrial, Scientific, and Medical}
\newacronym{dr}{DR}{Data Rate}
\newacronym{adr}{ADR}{Adaptive Data Rate}
\newacronym{ack}{ACK}{Acknowledgment}
\newacronym{tx}{Tx}{Transmitter}
\newacronym{rx}{Rx}{Receiver}
\newacronym{cr}{CR}{Coding Rate}
\newacronym{phdr}{PHDR}{Physical Header}
\newacronym{mhdr}{MHDR}{MAC Header}
\newacronym{mic}{MIC}{Message Integrity Code}
\newacronym{mtype}{MType}{Message Type}
\newacronym{fhdr}{FHDR}{Frame Header}
\newacronym{fport}{FPort}{Port Field}
\newacronym{frmpayload}{FRMPayload}{Frame Payload}
\newacronym{fctrl}{FCtrl}{Frame Control}
\newacronym{fcnt}{FCnt}{Frame Counter}
\newacronym{fopts}{FOpts}{Frame Option}
\newacronym{rfu}{RFU}{Reserverd For Future}
\newacronym{foptslen}{FOptsLen}{Frame Option Length}
\newacronym{fpending}{FPending}{Frame Pending}
\newacronym{fec}{FEC}{Forward Error Correction}
\newacronym{toa}{TOA}{Time on Air}
\newacronym{otaa}{OTAA}{Over-The-Air Activation}
\newacronym{aep}{AEP}{Activation By Personalization}
\newacronym{appeui}{AppEUI}{Application Identifier}
\newacronym{nwkskey}{NwkSKey}{Network Session Key}
\newacronym{appskey}{AppSKey}{Application Session Key}
\newacronym{nwkid}{NwkID}{Network Identifier}
\newacronym{nwkaddr}{NwkAddr}{Network Address}
\newacronym{lsb}{LSB}{Least Significant Bits}
\newacronym{msb}{MSB}{Most Significant Bits}
\newacronym{deveui}{DevEUI}{End-node Identifier}
\newacronym{appkey}{AppKey}{Application Key}
\newacronym{aes}{AES}{Advanced Encryption Standard}
\newacronym{itu}{ITU}{International Telecommunication Union}
\newacronym{etsi}{ETSI}{European Telecommunications Standard Institute}
\newacronym{erp}{ERP}{Effective Radiated Power}
\newacronym{lan}{LAN}{Local Area Network}
\newacronym{wan}{WAN}{Wide Area Network}
\newacronym{man}{MAN}{Metropolitan Area Network}
\newacronym{wap}{WAP}{Wireless Access Point}
\newacronym{pan}{PAN}{Personal Area Network}
\newacronym{pos}{POS}{Personal Operating Space}
\newacronym{afh}{AFH}{Adaptive Frequency Hopping}
\newacronym{ll}{LL}{Link Layer}
\newacronym{br}{BR}{Basic Rate}
\newacronym{edr}{EDR}{Enhanced Data Rate}
\newacronym{l2cap}{L2CAP}{Logical Link Control Adaptation Protocol}
\newacronym{gap}{GAP}{Generic Access Profile}
\newacronym{gatt}{GATT}{Generic Attribute Convention}
\newacronym{poi}{PoI}{Point of Interest}
\newacronym{mbps}{Mbps}{Megabits Per Second}
\newacronym{gbps}{Gbps}{Gigabits Per Second}
\newacronym{wifi}{Wi-Fi}{Wireless Fidelity}
\newacronym{wlan}{WLAN}{Wireless Local Area Network}
\newacronym{ieee}{IEEE}{Institute of Electrical and Electronics Engineers}
\newacronym{ssid}{SSID}{Service Set Identifier}
\newacronym{wep}{WEP}{Wired Equivalent Privacy}
\newacronym{wpa}{WPA}{Wi-Fi Protected Access}
\newacronym{tkip}{TKIP}{Temporal Key Integrity Protocol}
\newacronym{m2m}{M2M}{Machine to Machine}
\newacronym{afhs}{AFHS}{Adaptive Frequency Hopping Spread Spectrum}
\newacronym{plc}{PLC}{Power Line Communication}
\newacronym{ue}{UE}{User Equipment}
\newacronym{gsm}{GSM}{Global System for Mobile Communications}
\newacronym{gprs}{GPRS}{General Packet Radio Service}
\newacronym{3gsm}{3GSM}{Third generation Global System for Mobile}
\newacronym{cdma}{CDMA}{Code Division Multiple Access}
\newacronym{msc}{MSC}{Mobile Switching Center}
\newacronym{bts}{BTS}{Base Transceiver Station}
\newacronym{pstn}{PSTN}{Public Switched Telephone Network}
\newacronym{bsc}{BSC}{Base Station Controller }
\newacronym{lte}{LTE}{Long Term Evolution}
\newacronym{vlr}{VLR}{Visitor Location Register}
\newacronym{hlr}{HLR}{Home Location Register}
\newacronym{nbiot}{NB-IoT}{Narrow-band IoT}
\newacronym{dsss}{DSSS}{Direct Sequence Spread Spectrum}
\newacronym{prb}{PRB}{Physical Resource Block}
\newacronym{ec}{EC}{Extended Coverage}
\newacronym{3gpp}{3GPP}{Third Generation Partnership Project}
\newacronym{3g}{3G}{Third Generation}
\newacronym{sig}{SIG}{Special Interest Group}
\newacronym{unb}{UNB}{Ultra Narrow Band}
\newacronym{dbpsk}{DBPSK}{Differential Binary Phase Shift Keying}
\newacronym{tvws}{TVWS}{TV White Space}
\newacronym{gmsk}{GMSK}{Gaussian Minimum Shift Keying}
\newacronym{qpsk}{QPSK}{Quadrature Phase Shifting key}
\newacronym{rpma}{RPMA}{Random Phase Multiple Access}
\newacronym{gfsk}{GFSK}{Gaussian Frequency-Shift Keying}
\newacronym{gps}{GPS}{Global Positioning System}
\newacronym{fhss}{FHSS}{Frequency-Hopping Spread Spectrum}
\newacronym{udp}{UDP}{User Datagram Protocol}
\newacronym{ttn}{TTN}{The Things Network}
\newacronym{rssi}{RSSI}{Received Signal Strength Indication}
\newacronym{mqtt}{MQTT}{Message Queue Telemetry Transport}
\newacronym{json}{JSON}{JavaScript Object Notation}
\newacronym{per}{PER}{Packet Error Rate}
\newacronym{vswr}{VSWR}{Voltage Standing Wave Ratio}
\newacronym{ccm}{CCM}{Counter with CBC-MAC}
\newacronym{bfsk}{BFSK}{Binary Frequency Shift Keying}
\newacronym{amr}{AMR}{Automated Meter Reading}
\newacronym{ami}{AMI}{Advanced Metering Infrastructure}
\newacronym{han}{HAN}{Home Area Network}
\newacronym{qos}{QoS}{Quality of Service}

\usepackage{cite}
\usepackage{url}
\usepackage{booktabs}
\usepackage[super]{nth}
\usepackage{graphicx}
\usepackage{subfigure}
\usepackage{caption}
\usepackage{microtype}
\usepackage{balance}
\usepackage[utf8x]{inputenc} 
\usepackage{color}
\pdfoutput=1

\begin{document}
	
\newcommand{\lora}{LoRa\textsuperscript{\tiny{TM}}}
\newcommand{\lorawan}{LoRaWAN}
\newcommand{\mdot}{mDot}	

\title{Performance Evaluation of the \lora~Protocol in the context of Smart Meter}

\author{Muhammad Nouman~Rafi$^{1}$ and Muhammad Muaaz$^{2}$% <-this % stops a space
	\thanks{$^{1}$ Muhammad Nouman Rafi is a M.Sc. student at the Department of Energy Informatics of University of Applied Science, Hagenberg, Austria. 
		{\color{blue}{Email}}: nouman.rafi@gmail.com}%
	\thanks{$^{2}$Muhammad Muaaz is a researcher in  the Department of Information and Communication Technology of University of Agder, Norway.  
		{\color{blue}{Email}}: muhammadm@uia.no}%
}

\maketitle

\begin{abstract}
In recent years, the use of \gls{lpwan} is increasing for the \gls{iot} applications. 
In order to demonstrate the application of LPWAN technologies for a realistic smart metering scenario, we set-up and implement a widely used LPWAN protocol which is called \lorawan.~%to determine its aptitude for smart metering applications.
In this study, the \lorawan~is implemented by using Multitech devices (end-node and gateway) and the performance of the network is evaluated for different physical (e.g. location, distance etc.) and link parameters (e.g. data rate, transmission power etc.), under the European regulations. To evaluate the performance of the networks, we collected uplink packets in different indoor and outdoor scenarios at various locations. 
Our results show that \lorawan~is easy to setup, configurable, scalable, and it performs well  for real-time smart metering applications.
Moreover, it is necessary to evaluate the physical conditions for the selection of the available system parameters for deploying a robust \lorawan~network. 

\end{abstract}



\IEEEpeerreviewmaketitle

\section{Introduction}
Nowadays, in many countries, traditional electromechanical utility meters are being replaced with smart meters \cite{eu-27, 8523771}. A smart meter (whether it is for electricity, gas, or water) is an electronic device that records the consumption of the utility and reports it back to the utility provider in a secure fashion. Furthermore, smart meters allow \gls{amr} and \gls{ami} to offer a multitude of benefits to both consumers and utility providers \cite{van2006smart}. For instance, smart meters combined with in-home displays could provide consumers with real-time utility consumption information, thus giving them a better control over their utility spending and potentially contributing towards utility savings. Similarly, by analyzing the utility consumption behavior of consumers, utility providers can not only well manage the utility distribution network, but they can also monitor it in real time. Moreover, they are also cheap and report more accurate consumption \cite{MCKENNA2012807}. Due to these benefits, the roll-out of smart meters is increasing rapidly. Until 2020, the European Union (EU) expects that almost 72\% of the European consumers will have smart meters for electricity and 40\% will have smart gas meters \cite{smrollout}. Counties in EU, like Austria, will have 95\% of the smart meters rolled-out by 2020 \cite{smrollout}.  

For smart metering connectivity, cellular technologies are widely used in Europe and they fulfill the metering system requirements \cite{erlinghagen2015smart}. But these technologies are costly and have limited battery life. Whereas,  other connectivity options for smart metering are \gls{plc}, Ethernet, Wi-Fi, and Bluetooth. However, these technologies don't offer long range connectivity \cite{background}. Therefore, to meet the massive target of smart meter roll out and to deal with the connectivity issues, utility providers need to find a better alternative which is more reliable and cost effective.

Recently, a new wireless communication technology called \gls{lpwan} has emerged. \gls{lpwan} describes a class of wireless communication technologies which are designed for long range communication between things (connected objects) at the cost of low bit rate.  Therefore, it fulfills the requirements for smart metering applications \cite{8523771}.
Whereas, the traditional technology networks (e.g. Cellular and \gls{plc}) until now are still less favorable for transmitting short messages occasionally over a long range in a cost effective manner \cite{varsier2017capacity}. Fortunately, the AMI for smart meters allows integrating the \gls{lpwan} technologies. Over the last five years, several proprietary and open standard  \gls{lpwan}  technologies have been introduced in licensed and unlicensed frequency bandwidth. Among the available \gls{lpwan} technologies, \lora,~Sigfox, and NB-IoT are the market leaders and widely used. \lora~and Sigfox are better in terms of cost, network capacity, and battery life-time. Meanwhile, NB-IoT offers better Quality of Service (QoS) and low latency \cite{MEKKI2018}. Since its inception, \lora~has seen an exponential growth even though the technology is still in its beginning, it is the most adapted technology for the \gls{iot} applications \cite{adelantado2017understanding}.  
Therefore, in this article, we will evaluate the performance of the \lora~protocol under different conditions in the context of smart meter by setting up a \lorawan~with real hardware under European regulations. The results obtained from this study can be used for better integration of the technology for smart metering applications. 

Rest of the article is organized as follows, section~\ref{relatedwork} presents a brief overview of various other studies in which researchers have evaluated the \lora~protocol for different applications. A detailed overview of the \lora~protocol, \lorawan~architecture, \lorawan~end node classes, \lorawan~security, and the European regulations are given in section~\ref{lorawan}. An overview of our \lorawan~setup and application to generate and transmit smart metering data is given in section \ref{systemSetup}. Results of the indoor and outdoor evaluations of the \lora~protocol are given in sections~\ref{results} and \ref{outdoor}, respectively. Finally, the conclusion of this study is given in section \ref{conclusion}.

\section{Related Work}
\label{relatedwork}
There exist various parameters in a \lorawan, which need to be taken into consideration while evaluating the performance of \lorawan. For instance, in \cite{aref2014free}, authors studied the outdoor range of the \lorawan.~The experiment was performed in Germany with 250 KHz bandwidth and they recorded \gls{per}, \gls{snr}, and \gls{rssi} for different distances with 10, 50 and 100 bytes of payload length. Obtained results showed that with a payload of 10 bytes, the packets are successfully delivered with zero \gls{per} up to 8 km. However, the \gls{per} increases with the increase of the payload. The payload of 50 bytes resulted in invalid packets from the range of 2.3 km and with a payload of 100 bytes, the \gls{per} was near zero for up to 6 km. This study was conducted by using 250 KHz bandwidth for channels, whereas, \lorawan~typically uses 125 KHz bandwidth for the channels.

In \cite{tanumihardja2015application}, the troughs water level is monitored by using \lorawan~with 915 MHz band. Their outdoor experiment results show that the location of the end-nodes has a major impact on the network performance. Additionally, the quality of transmission is poor, when end-nodes are placed closer to the ground. Their simulated results of 100 nodes networks showed that increasing the \gls{dr} of 5 nodes within the network, the delivery of valid packets drops by 17\%. The experiment was done in the \gls{ism} band of the USA, which does not have the same regulation like the \gls{ism} band of EU (in USA band, there is no duty cycle limitation).

Wixted et al. \cite{wixted2016evaluation} performed an outdoor evaluation of \lorawan~and \lora~using three gateways in Glasgow. Outdoor testing documented \gls{rssi}, GPS location, and reliability of receiving the \gls{ack} after transmitting. Reliability testing showed that cellular connections of gateways experience disconnected periods, because of inactive policies of the cellular networks. Continuously pinging the gateways increased the connectivity rate of \lorawan~from 70 to 95\%. %The data packet was received in 2.5\% attempts, but the \lora~end-node didn’t receive an acknowledgment.
A coverage range test was also conducted with two \lora~transceivers. In one direction, connectivity test was successful till 2.2 km and in the other direction, the range was 1.6 km due to a hill between the end-node and gateway. \cite{wixted2016evaluation}.

Another study \cite{petajajarvi2015coverage} was conducted in Finland in order to check the \lora~range capabilities by using a mobile end-node (boat and car). The \gls{rssi} values, packet loss, and GPS location were recorded up-to 30 Km. The end-node had a \gls{dr} 0, while transmission power was fixed to 14 dBm.
The measurement results of the car showed that within the range of 2 km from the gateway or base station, the \gls{rssi} value was mostly greater than -100 dBm and 12\% transmitted packets were lost. The packets lost were increased with the increase of distance from the gateway. whereas in the boat, 15\% of the packets were lost within the range of 2--5 km. Lastly, path loss exponents were also measured for the car and boat. 

In \cite{centenaro2016long}, the authors evaluated the coverage range of \lora~network for different \gls{dr}s. The gateway was located at the top of a tall building (19 floors) with antenna gain 0. The packets ware received up-to 2 km, but the authors felt that the range should be assumed 1.2 km due to variations in the link budget.

The researchers of \cite{augustin2016study} tested the network range of \lora~in the urban area of Paris, France. During the test, transmission power was fixed to 14 dBm and vary the \gls{dr} 0, 3 and 5. The maximum range was achieved with a \gls{dr} 0 up-to 3.4 km and 38\% packets were delivered. However, the closest location (650 m) of the test revealed that with a \gls{dr} 0, 100\% of the packets can be delivered and this value drops to 84\% when a \gls{dr} 5 is used. During the test, the \gls{ack} was disabled.

An indoor \lora~network deployment evaluates the indoor performance of \lorawan \cite{neumann2016indoor}. Tests results show that due to EU limitation of the duty cycle, with \gls{dr} 0 and with a maximum size of the packet, an end-node should wait for approximately 4 minutes and 30 sec before generating a new transmission. Even with no data in \gls{dr} 0, an end-node waits for 2 minutes between consecutive transmissions. On the other side, with \gls{dr} 5 or \gls{sf} 7, the delay is minimized to 2 seconds. Also, their results show that \gls{rssi} values are not changed by changing the \gls{dr}, but the \gls{snr} values are changed. The transmission test was repeated with a \gls{dr} 2 and found that if the end-nodes are near to the gateway, then packets duplications and bad \gls{crc} packets were increased \cite{neumann2016indoor}. 

Another study \cite{gregora2016indoor} was conducted in Prague for an indoor penetration of \lorawan~signal. In this study, an end-node was placed in different locations and floors inside the building and recorded the end-nodes \gls{rssi} values. The gateway was placed on the rooftop. For each location, 10 packets were sent with a larger payload, which contains information of \gls{snr} and \gls{rssi} etc. Tests were performed with  IMST iU880A \lora~node, which does not have an external antenna feature. The power was fixed at 20 dBm. The authors concluded the result from recorded \gls{rssi} that the gateway placing on the rooftop provides more coverage range as compare to the basements.

In \cite{augustin2016study}, the authors study the signal strength levels 
for message packets, which were transmitted from an end-node to the gateway. The end-node was located outdoors and the gateway was located indoors. In this experiment, the transmission power was fixed at 2 dBm. The results of the experiment show that the maximum coverage range is 100 m. while \gls{rssi} didn't decrease with the increase of \gls{sf}. It was also noted that decrease in the \gls{rssi} value had increased the \gls{snr}.

The authors analyzed the \lorawan~capacity limits for smart metering applications using the network simulator for both the uplink communication and downlink communication in \cite{varsier2017capacity}. The results showed that, for a case where smart meters are located deep indoor, the network covering 17 km of an area with 19 gateways, which will be located 1 km from each other. It could be sufficient for achieving 98\% quality of service if considering only uplink communication. On the other hand, if both the uplink communication and downlink communication are taken into consideration, the network capacity will dramatically decrease. 

\begin{figure*}[t]
	\centering
	\includegraphics[width=0.80\linewidth]{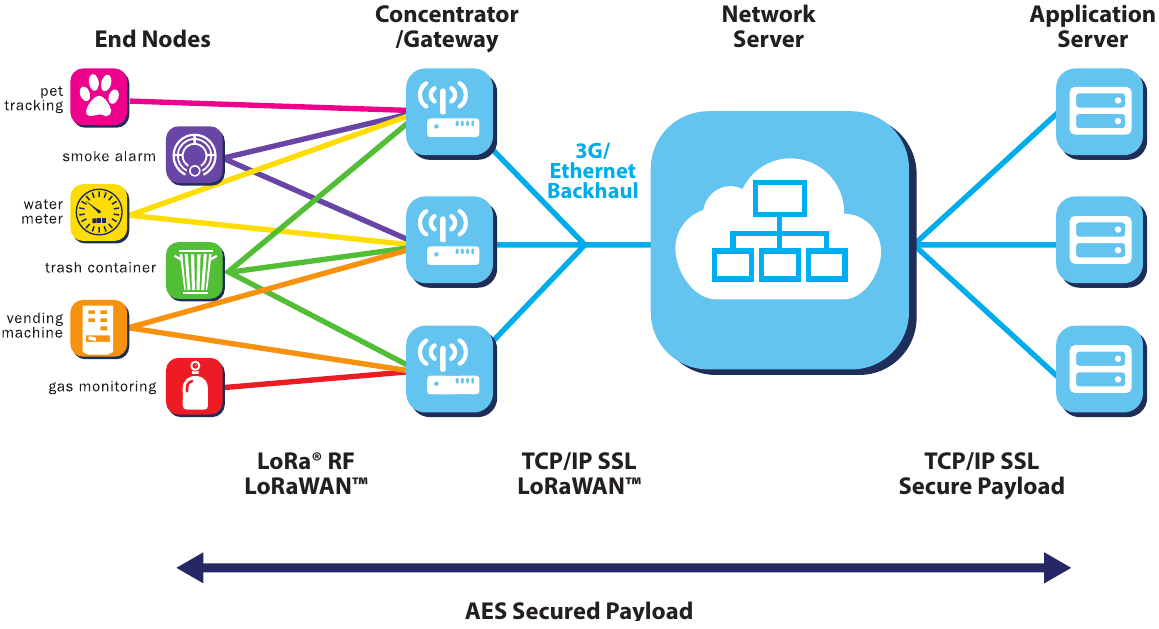}
	\caption[\lorawan~network architecture]{\lorawan~network architecture~\protect\cite{technical_overview_LoRa}.}
	\label{ch3_f_3}
\end{figure*}

\section{LoRa and LoRaWAN}
\label{lorawan}
\lora~is the \gls{phy} protocol designed and implemented by Semtech\footnote{\url{www.semtech.com}}. It operates in unlicensed sub-GHz \gls{ism} bands (i.e. 868 MHz in Europe, 915 MHz in North America, and 433 MHz in Asia.), and it is based on \gls{css} modulation technique \cite{technical_overview_LoRa, css}. Like the \gls{fsk} modulation technique, the \gls{css} modulation technique maintains the same low power characteristics, but it has an increased wireless communication range \cite{sx1272}. 

There exist various parameters in the \lora~modulation technique like spreading factors (SF7 to SF12), channel bandwidth, and coding rate (CR). These parameters can be used to adapt the data rate and range trade off. For instance, a low spreading factor allows a short range communication link but  offers high data rate, whereas, a high spreading factor offers a long range communication link at the cost of low data rate. The data rate of the \lora protocol is between 300 bps and 5000 bps. Further, messages transmitted using different spreading factors are practically orthogonal to each other, therefore, they do not affect each other, thus they can be received simultaneously by a \lora~gateway \cite{technical_overview_LoRa, sx1272}. 

The main advantage of \lora~is long range. A base station or the single gateway can cover several kilometers of distance. The range highly depends upon the obstructions or environment of the location, but \lora~offers a greater link budget compared to other wireless communication technologies. The maximum link budget of \lora~is 157 \gls{db} \cite{sx1272}.

In 2015, a \lora~protocol based wireless communication network called ``Long Range Wide Area Network'' \lorawan~was created by the \lora~alliance in cooperation with IBM, Semtech, Microchip, and Actility \cite{LoRaWAN_Specification}. \lorawan~defines the wireless communication protocol and the architecture of the network. The architecture of  the network and the protocol have the most impact on determining the capacity of the network, battery life of the node, service quality, and security \cite{technical_overview_LoRa}.

\subsection{\lorawan~Architecture}
The overall \lorawan~consists of 4 blocks: an end-node, a gateway,  a network server, and an application server, as shown in Fig.~\ref{ch3_f_3}. \lorawan~is normally laid out in a star topology in which end-nodes are connected with a single or multiple gateways via the single-hop \lora~link \cite{LoRaWAN_Specification}. 

End-nodes form the base of the \lorawan,~they are low power sensors or actuators that send a small amount of data to the network server via a concentrator/gateway. A gateway is connected to a public or private network server via standard \gls{ip} (either Ethernet, \gls{wifi}, satellite or cellular). In \lorawan,~end-nodes or motes are not required to send data to a particular gateway, but the end-node broadcasts its data to every gateway within the network. The gateway acts as a bridge and sends packets to the network server by adding additional information about the quality of the packets \cite{technical_overview_LoRa}. 

The network server filters unwanted and duplicate packets and replies to the end-nodes through certain in-range gateways. Communication in the \lora~network is bi-directional.  Communication from an end-node to the network server via a gateway is called uplink communication and its vise versa is called downlink communication. The network server has information for every node in the entire network, so it knows specifically where to send the downlink packets. The network server manages the whole network and the \gls{adr} mode can automatically manage the \gls{sf} for all end-nodes with the \gls{adr} scheme \cite{ruano2016lora, centenaro2016long}. 
Finally, the application server is responsible for the end-node “inventory” part of the \lorawan,~handling join-requests and decryption of application payloads.

\subsection{\lorawan~End-node Classes}
\label{classes}
\lorawan~defines three different classes of end-nodes (i.e. Class A, Class B, and Class C) to support a variety of \gls{iot} applications \cite{MEKKI2018}. All these classes of end-nodes allow bidirectional communication, however, the basic difference is in their downlink communication. 

For instance, in Class A, every uplink transmission is pursued by two short time receive windows for the downlink communication \cite{LoRaWAN_Specification}. A class A type end-node is used for low power actuators
or sensors without latency constraint. Mostly, this type of end-nodes are used to measure
the temperature, the metering data, and for the tracking etc. 

A Class B type end-node opens extra receive windows for the downlink communication compared to a Class A type end-node. Receive windows in Class B end-nodes are opened at scheduled times \cite {LoRaWAN_Specification}. In order to schedule a receive window to open, the end-node receives a time synchronized beacon from the gateway, allowing the network server to determine when the end-node is listening. These type of end-nodes can be most
useful for battery devices like controlled reading with sensors and alarm sensors etc. 

Class C type of end-nodes continuously listen for
downlink transmission after uplink transmission \cite{LoRaWAN_Specification}. End-node of class C consumes
more power than class A and class B, but the latency rate is lower for communication
between the server and the end-node.

\subsection{\lorawan~Network Security} 

Security is a major prerequisite in every wireless communication network. The same security prerequisite has also been planned for the \lorawan~network. To personalize the end nodes in \lorawan~network, every end-node is customized with an \gls{aes}-128 bit \gls{appkey} and \gls{deveui} based on IEEE EUI-64. The \gls{appkey} and \gls{deveui} are obtained from the network server \cite{LoRaWAN_SECURITY}. %In any case, a network of \lorawan~is recognized by utilizing a 24-bits unique identifier \cite{LoRaWAN_Specification, technical_overview_LoRa}. 

Numerous properties are strengthened to guarantee the \lorawan~security, for example, integrity protection and mutual authentication. Mutual authentication is set up in the network security section between end-node and the network server. This authentication guarantees that only recognized end-node can join the network \cite{LoRaWAN_Specification}. Consequently, two security keys are determined, the first \gls{nwkskey} is used for \gls{mac} protection and the second \gls{appskey} is used for end-to-end encryption of the payload. \gls{nwkskey} is shared in the network which is in between the node and the network server in order to authenticate \lora~message packet. Whereas, \gls{appskey}  is shared in the application layer between the node and the application server to encrypt and decrypt the application payload~\cite{LoRaWAN_SECURITY, LoRaWAN_SECURITY_faqs}. 

\begin{figure}[h]
	\centering
	\includegraphics[width=1.0\linewidth]{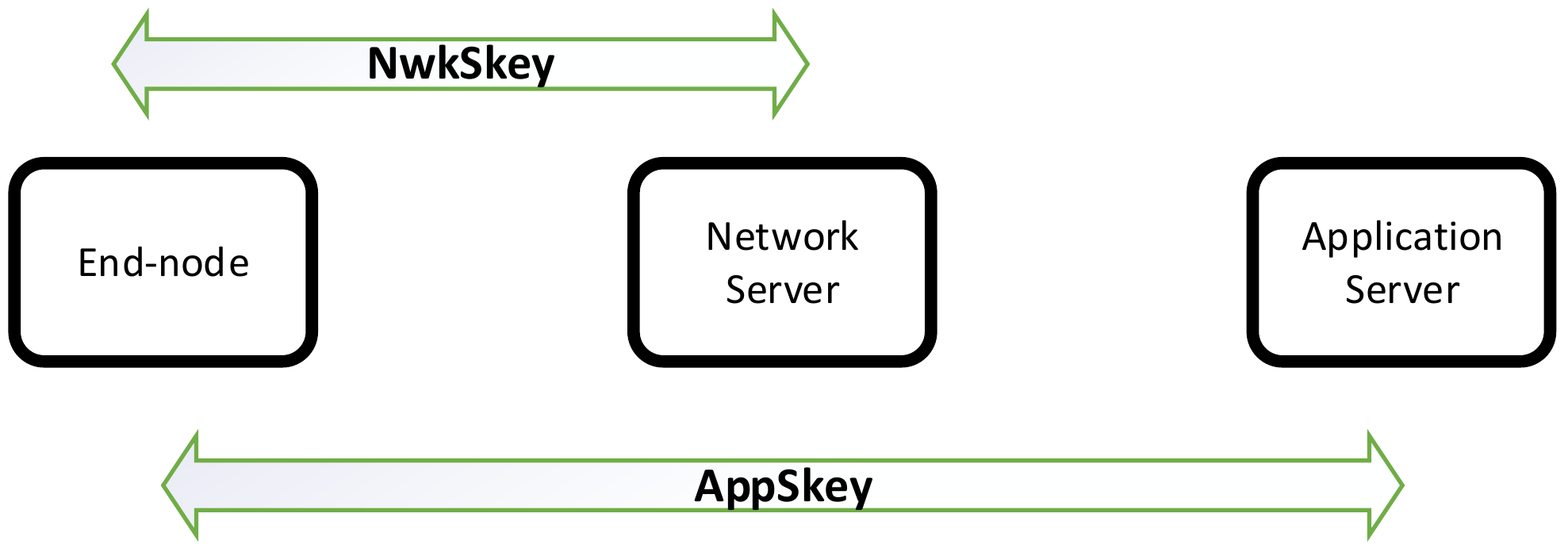}
	\caption[Authentication and encryption in \lora~network]{Authentication and encryption in \lora~network.}
	\label{ch3_f_17}
\end{figure}

Fig.~\ref{ch3_f_17} demonstrates the two keys that are used for authentication and encryption in the network server and the application server \cite{LoRaWAN_SECURITY}. Both session keys (\gls{nwkskey} and \gls{appskey} are unique for every end-node, and for every session. End-nodes can either join the network dynamically (Over the Air Activation (OTAA)) or statically (Activation by Personalization (ABP)) \cite{LoRaWAN_SECURITY}. In the case of dynamic activation, both session keys are regenerated every time an end node joins the network by using the \gls{appkey}. In the case of static activation, both session keys stay the same until one manually changes them. 

\begin{figure*}[!t]
	\centering
	\includegraphics[width=1.0\textwidth]{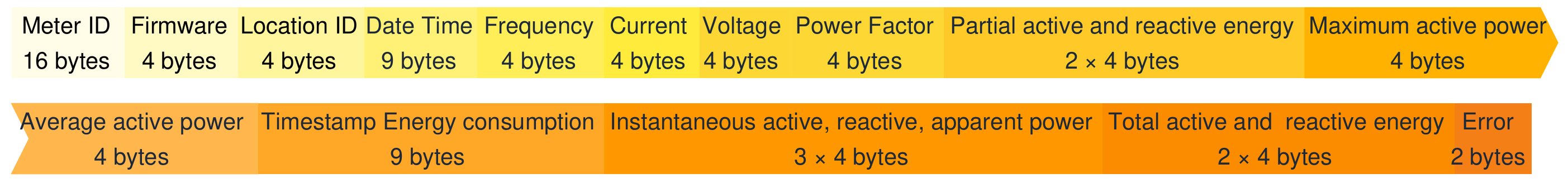}
	\caption[A 96 bytes payload.]{A 96 bytes smart meter datagram for SFs 6, 7, and 8.}
	\label{96bytes}
\end{figure*}

\begin{figure*}[!t]
	\centering
	\subfigure[First packet]
	{
		\includegraphics[width=\textwidth]{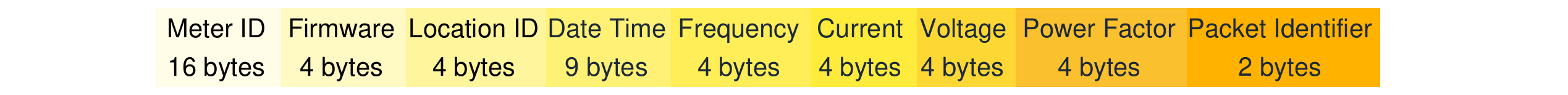}
		\label{fig:first_sub}
	}
	\subfigure[Second packet]
	{
		\includegraphics[width=\textwidth]{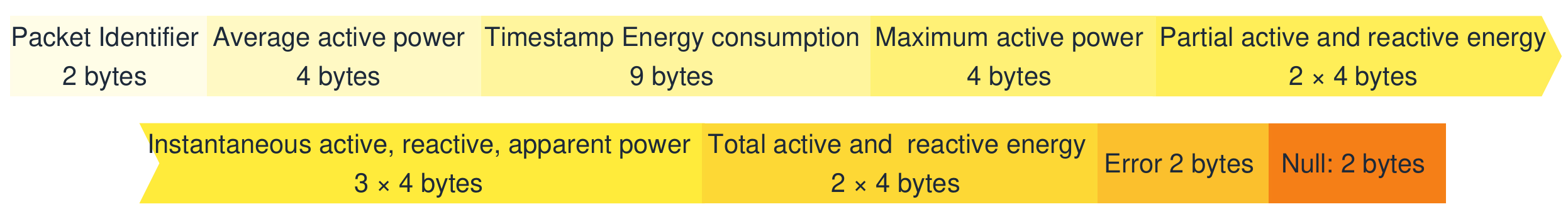}
		\label{fig:second_sub}
	}
	\caption{Two 51 bytes smart meter datagrams for SFs 9, 10, and 11.}
	\label{51bytes}
\end{figure*}
\subsection{European Regulation}

Although the \lora~protocol operates in unlicensed frequency bandwidth,
there are some regulations to use the \lora~network. These regulations may vary from region to region and are set at the following three levels \cite{LoRaWAN_Regional_Parameters}:
\begin{enumerate}
	\item at the national level,
	\item at the European (EU) level, which is set by the European Commission,
	\item at~the~global~level, which is set~by~the~\gls{itu}.\\
\end{enumerate}
For instance, in the EU region, the \lorawan~operates in 863-870 MHz unlicensed frequency band and according to the \gls{etsi}, 
three sub-GHz channels 868.1 MHz, 868.3 and 868.5 must be used in \lorawan~\cite{ETSI_Compliance_SX1272, ETSI}. The duty cycle and \gls{erp} limitation are shown in Table~\ref{ch3_t_7} \cite{ETSI,CEPT}.

\begin{table}[h]
	\centering
	\begin{tabular}{lccc}
		\toprule
		\textbf{Frequency} & \textbf{BW} & \textbf{Duty Cycle} & \textbf{Max ERP} \\
		\midrule
		868.1 & 125 kHz & 1\% & 14 dBm \\ \hline
		868.3 & 125 kHz & 1\% & 14 dBm \\ \hline
		868.5 & 125 kHz & 1\% & 14 dBm \\ \hline
		868.85 & 125 kHz & 0.1\% & 14 dBm \\ \hline
		869.05 & 125 kHz & 0.1\% & 14 dBm \\ \hline
		869.525 & 125 kHz & 10\% & 27 dBm \\ 
		\bottomrule
	\end{tabular}
	\caption[Limitation of \lorawan~channels]{Limitation of \lorawan~channels~\protect\cite{ETSI_Compliance_SX1272, LoRaWAN_Regional_Parameters}.}
	\label{ch3_t_7}
\end{table}

\section{System Setup}
\label{systemSetup}
This study is aimed to evaluate the performance of the \lora~protocol for smart metering application under the guidance of European regulations using real hardware. Therefore, we explain the hardware and software details of our \lorawan~setup.

\begin{figure}[h]
	\centering
	\includegraphics[width=1.0\linewidth]{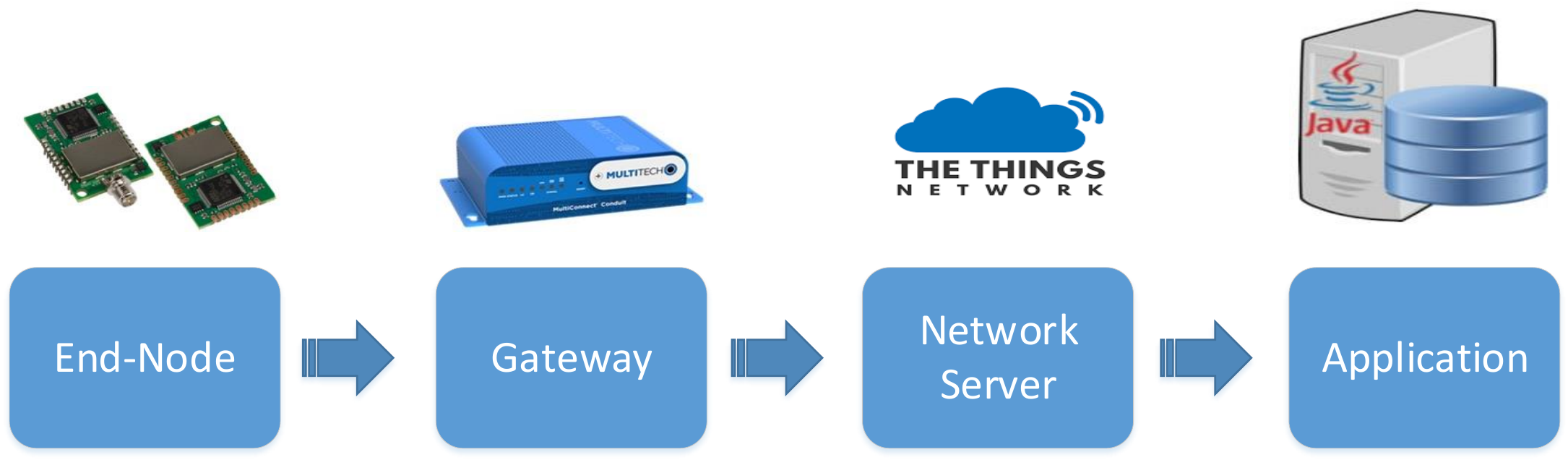}
	\caption[Block diagram of our \lorawan~setup]{Block diagram of our \lorawan~setup.}%~\lora~\protect\cite{LoRaWAN_Specification}.}
	\label{ch4_f_1}
\end{figure}

The architecture of our \lorawan~consists of a MultiConnect \mdot\footnote{https://www.multitech.com/brands/multiconnect-mdot}~(end-node), a MultiConnect conduit gateway\footnote{\url{https://www.multitech.com/brands/multiconnect-conduit}} , an open source and public network called ``\gls{ttn}\footnote{\url{www.thethingsnetwork.org}}''. \gls{ttn} gives us a network server as well as all the components which are required for coupling all the devices \cite{thethingsnetwork_Online}, as shown in Fig.~\ref{ch4_f_1}.
In the first step, both end-node and gateway are configured to connect and communicate with each through the \lora~physical layer. After that, both end-node and gateway are set up with \gls{ttn} console. 

To capture the uplink packets, a Node.js application is created with the help of Node.js Application SDK for the \gls{ttn}. This applications is nothing but a logger. It connects to the \gls{ttn} by using the \gls{ttn} application access key and after connecting it stores uplink packets with their metadata in a local database in \gls{json}\footnote{\url{www.json.org/}} format.

\subsection{Application for end-node to generate the dummy data}
\label{applicaiton}
In our \lorawan~setup, the end-node (\mdot)~acts as a smart meter. The smart metering application requires 96 bytes to transmit the data as shown in Fig. \ref{96bytes}. For testing a smart metering scenario, a C++ application is developed for end-node by using Multitech libraries\footnote{\url{os.mbed.com/platforms/MTS-mDot-F411/}}. The end-node uses this application for transmitting data packets to the gateway and \gls{ttn} application. The end-node periodically transmits the \lora~physical payload frames of 109 bytes for \gls{sf} 7, 8, 9 (\gls{dr} 5, 4, 3) and 64 bytes  for \gls{sf} 10, 11, and 12 (\gls{dr} 2, 1, 0), respectively. 

For \gls{sf} 7, 8, and 9, a 96 bytes dummy payload is used whereas a 51 bytes dummy payload is used for the last three \gls{sf}s, which is the maximum allowed payload for \gls{sf}s 10, 11, and 12. Therefore, for the last three \gls{sf}s, the end-node application divides the 96 bytes data packet into two 51 bytes data packets as shown in Fig.~\ref{51bytes} and sequentially transmits them. The \gls{ttn} application combines these two 51 bytes packets to a single 96 bytes packet by using the packet identifier field once received by the \gls{ttn} network server.  The remaining 13 bytes for all \gls{sf}s were for other fields of physical payload in the frame. These fields are \gls{mhdr} (1 byte), \gls{mac} Payload (8 bytes) and a \gls{mic} (4 bytes). %The details of these fields are explained in section \ref{frames} \cite{LoRaWAN_Specification}. 
%Since these tests were performed in Europe, and there is a 1\% limitation of the duty cycle.%, thus we set the delay time of 18 ms between 2 consecutive transmissions. %For each location, we collected 1 hour of data for each \gls{sf}.

\section{Results of the Indoor evaluation of the \lora~protocol}
\label{results}

\begin{figure*}[h]
	\centering
	\includegraphics[width=1.0\linewidth]{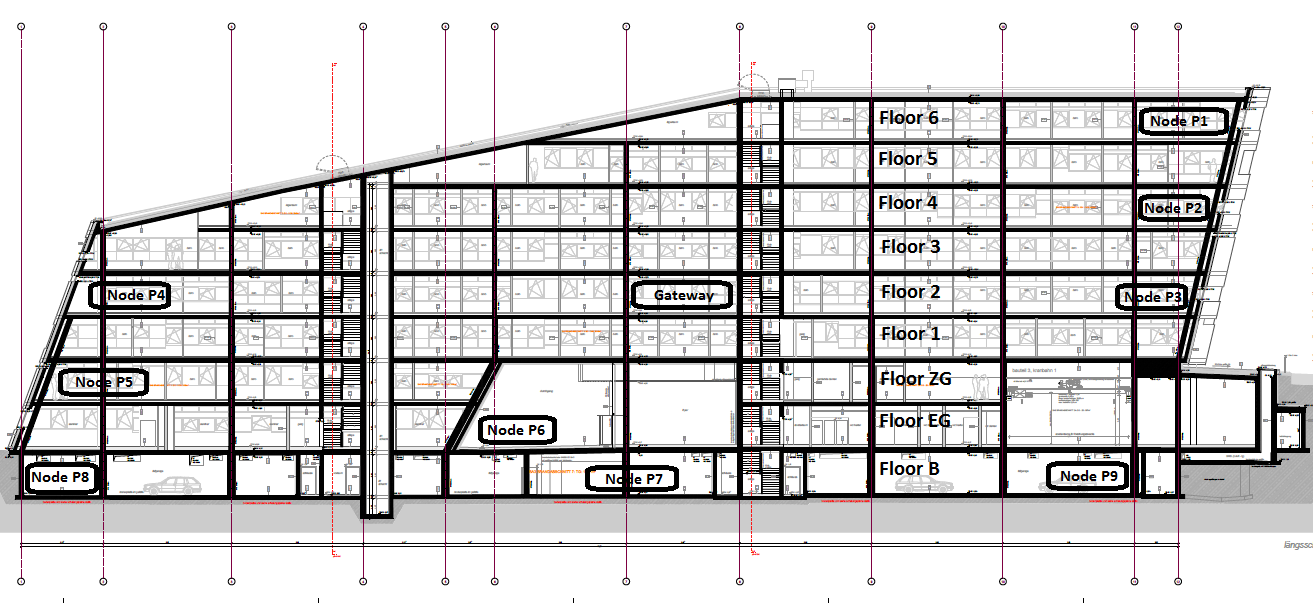}
	\caption[Indoor location of the gateway and different end-node positions]{Indoor location of the gateway and different end-node positions.}
	\label{ch5_f_5}
\end{figure*}

For indoor testing, we selected the Science Park 3 building, located in the Johannes Kepler University\footnote{\url{www.jku.at}} of Linz, Austria. %, because this building was suitable for indoor testing in the terms of length, width and floors. 
The infrastructure of this building is made of steel and hard concrete, and it consists of 9 floors including the basement. %The shape of the building is like a cruise. 
The dimensions of the building are as follows, the length is 84 m, width is 21.5 m including walls and the height is 27.5 m including the basement. We are using MultiTech products. MultiTech claims deep penetration of signal inside the building and we are using an isotropic antenna. So, we placed our MultiTech  gateway in the middle of the building on the \nth{2} floor and selected 9 different positions in the building at different floors and locations. The location of the gateway and different positions, where end-node was placed inside the building are shown in Fig.~\ref{ch5_f_5}. 

\begin{figure}[h]
	\centering
	\includegraphics[width=1.0\linewidth]{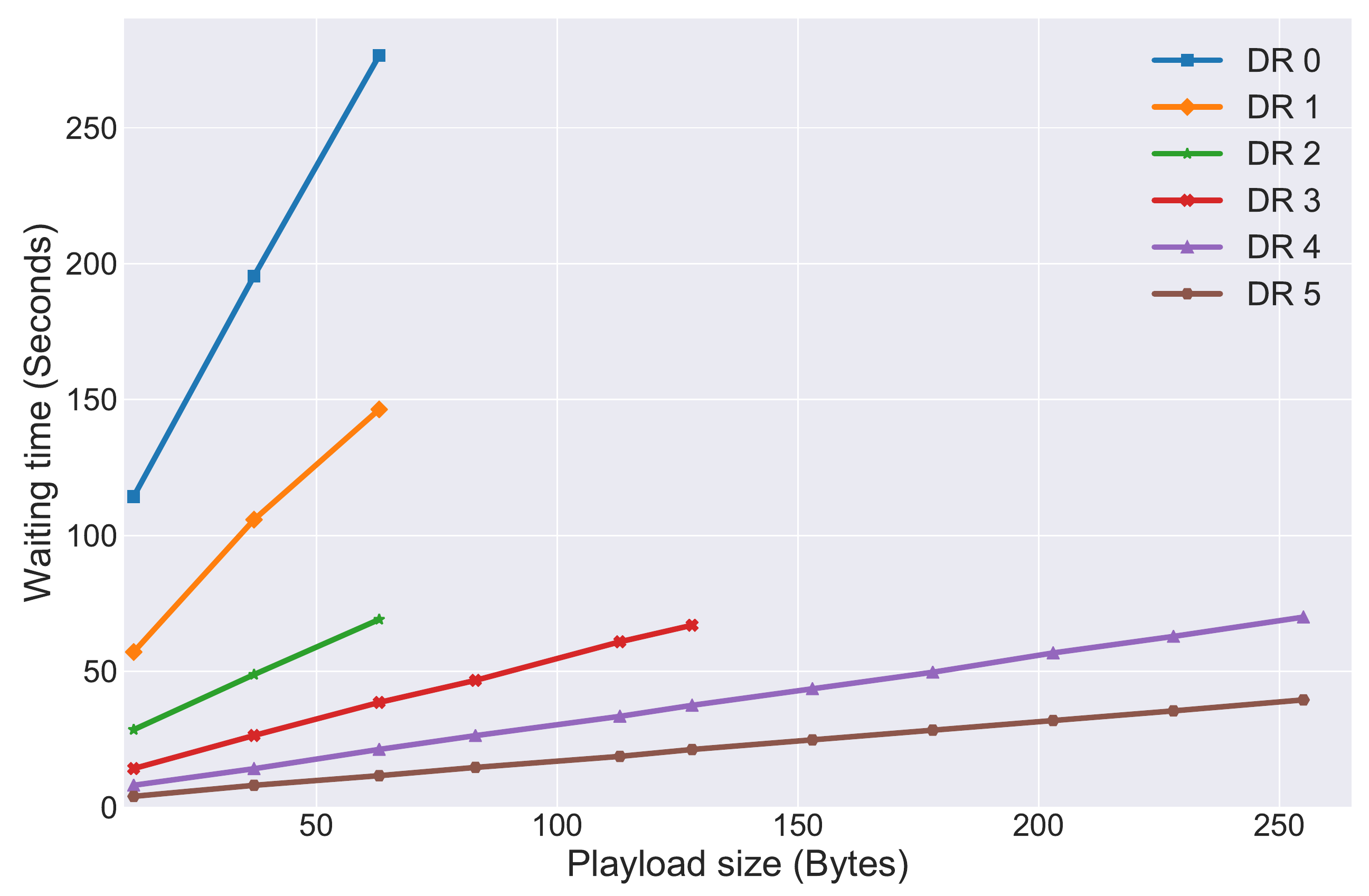}
	\caption[Waiting time for different payloads at 1\% duty cycle]{Wait time for different payloads at 1\% duty cycle.}
	\label{ch5_f_7}
\end{figure}
\begin{figure}[h]
	\centering
	\includegraphics[width=1.0\linewidth]{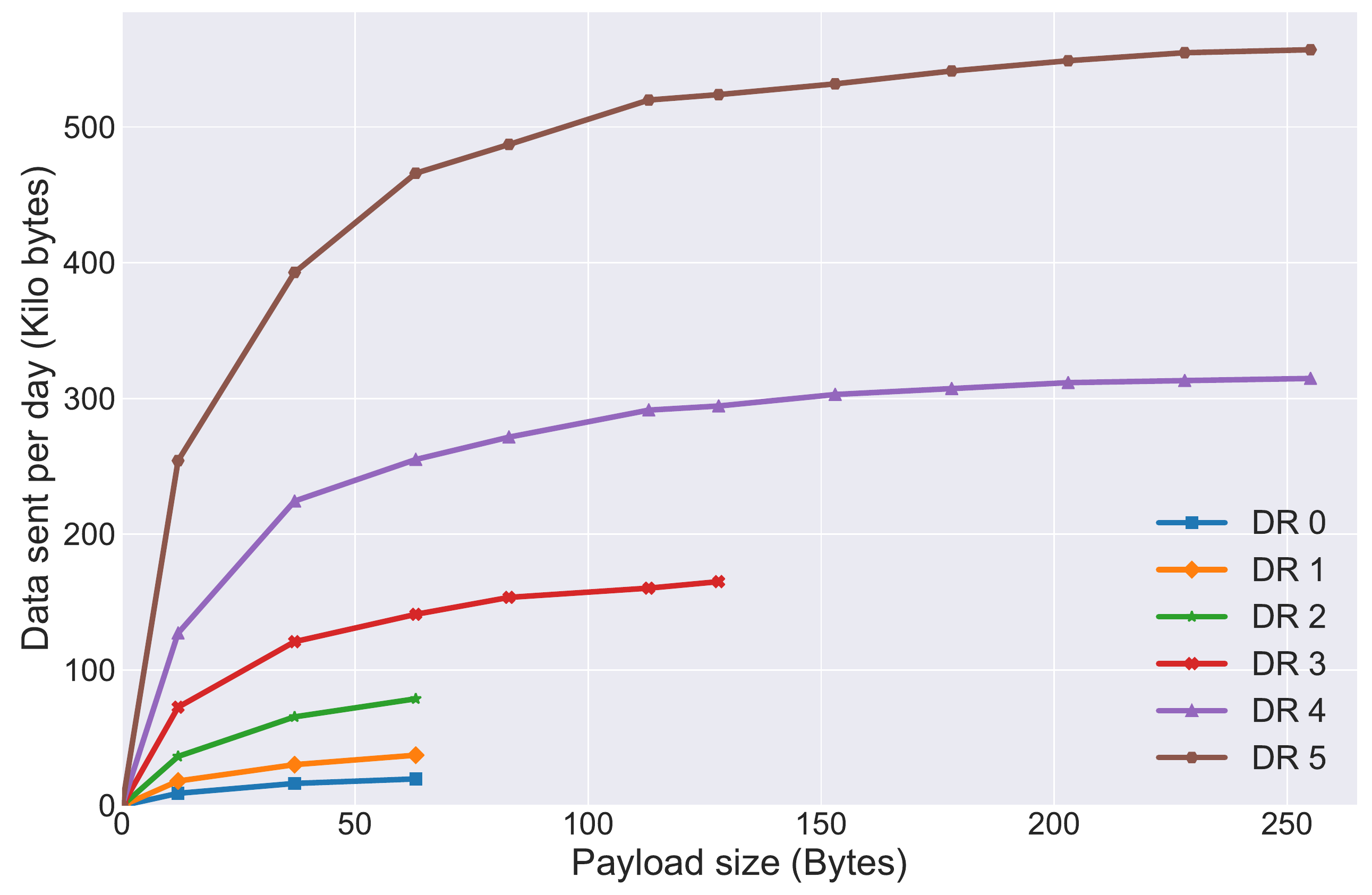}
	\caption[Total data sent per day for different DR and payloads  at 1\% duty cycle]{Total data sent per day for different DR and payloads at 1\% duty cycle.}
	\label{ch5_f_8}
\end{figure}
At each position, we recorded data for 6 hours (1-hour for each \gls{sf} (\gls{sf}7--\gls{sf}12)). Altogether, we recorded data for 54 hours for our indoor scenario. 
Moreover, we have also performed the following tests for our indoor scenario. These tests were performed at different time and days. 

\begin{itemize}
	\item Limitation of Sub-Band/channel Utilization.
	\item Received Signal Strength Indication (RSSI).
	\item Signal-to-Noise Ratio (SNR).
	\item Packet Error Rate (PER).
	\item \lora~network capacity.
	\item Impact of different \lorawan~classes on Acknowledgment (ACK).
\end{itemize}

\subsection{Limitation of Sub-Band/Channel Utilization}

Due to the limitation of sub-band/channel utilization of 1\% from the European regulation, the end-node must wait for a sufficient amount of time between two consecutive transmissions. It means that the end-node can only use 1\% of the total time to transmit data on the sub-band. Different \gls{dr}s and frame sizes can take different \gls{toa} for transmission.

Fig.~\ref{ch5_f_7} shows the wait time after packet transmission for each \gls{dr} and payload size. %The \gls{phy} payload consists of \gls{mac} header, \gls{mac} payload, and a \gls{mic}. This adds extra 13 bytes to the frame if no other fields are used.

From Fig.~\ref{ch5_f_7}, it can be observed that with \gls{dr} 0 or \gls{sf} 12 and with a maximum size of the packet, an end-node should wait for approximately 4 minutes and 30 sec before generating new transmission. Even with no data in \gls{dr} 0, the end-node waits for 2 minutes between consecutive transmissions. On the other side, with \gls{dr} 5 or \gls{sf} 7, the delay is minimized to 2 seconds. %Whereas in \cite{neumann2016indoor, blenn2017lorawan}, the authors also finds similar results about the limitation of the network. However our goal is to check the effect of the limitation of \lora~network in the context of smart metering. 
According to the application needs of smart metering, the \gls{dr} should be selected carefully in order to increase the scalability of the network.

Fig.~\ref{ch5_f_8} illustrates the maximum amount of data an end-node can send per day over a single channel with 1\% duty cycle. We can see that the amount of data in \gls{dr} 0 is limited. However, in \gls{dr} 5,  we can exchange data up to 550 kilobytes per day. As an outcome, \gls{dr} 5 seems more suitable for sensitive applications. After all, the different \gls{dr}s help to decrease collision between various end-nodes. %In \cite{blenn2017lorawan}, the Blenn et al.  also presents estimated results theoretically for data limitation send per day for different \gls{dr} and payloads at 1\% duty cycle, which are quite similar to our practically results. 
Fig. \ref{ch5_f_7} and \ref{ch5_f_8} represent the limitation of the \lora~network. The performance of the \lora~network can be increased by increasing the number of channels or increasing the duty cycle. 

\subsection{Received Signal Strength Indication (RSSI)}

\gls{rssi} is a very important parameter for wireless communication. This  parameter indicates the signal strength received by the antenna after cable losses in wireless communication. %The unit used to measure \gls{rssi} is dBm. dBm is  the ratio of the power in decibels of the total power in response to one milliwatt (mW). 
Since \gls{rssi} indicates the level of power, the signal is the strongest when the value of \gls{rssi} is the highest. Usually, the \gls{rssi} value is represented in negative form so an \gls{rssi} value close to 0 indicates better signal strength. \gls{rssi} normally depends on the distance, transmission power, and antenna gain \cite{sauter2010gsm}. %In all of the below figures, the \gls{rssi} is always represented on the y-axis. 
We performed various tests in order to evaluate the behavior of \gls{rssi} in the entire network. For each test, one hour of readings were acquired of the parameters being tested. %As expected, the number of packets received by the gateway increased with the increase in \gls{dr}.

\subsubsection{Impact on RSSI by changing the transmission power}
\label{2}
\begin{figure}[t]
	\centering
	\includegraphics[width=1.0\linewidth]{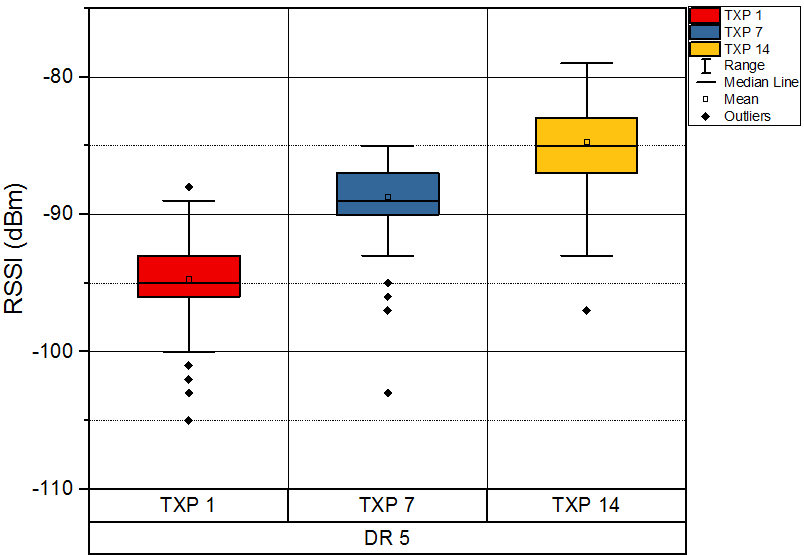}
	\caption[RSSI values by changing the transmission power]{RSSI values by changing the transmission power.}
	\label{ch5_f_12}
\end{figure}

The end-node (mDot) allows a range of transmission power from 1 dBm to 30 dBm. However, European regulation allows maximum transmission power up to 14 dB in the \lora~network. Therefore, for testing purposes, the range from 1 dBm to 14 dBm was used. This test was performed with a fixed \gls{dr} of \gls{dr} 5 and the gain was fixed to 3 dBi. All readings were taken at the same location. The x-axis represents the transmission  power. Fig.~\ref{ch5_f_12} shows the behavior of \gls{rssi} with the change in transmission power. The colored portion in each box represents 25\% to 75\% of values of \gls{rssi}. The outside whiskers show the complete range of the values. The mean value is represented by the clear square and the median is represented inside each box as a thick line. The outlying values are shown by the diamond shape.

A clear trend can be observed that the signal strength increases with the increase in transmission power.% \cite{cattani2017experimental}.

\begin{figure}[t]
	\centering
	\includegraphics[width=1.0\linewidth]{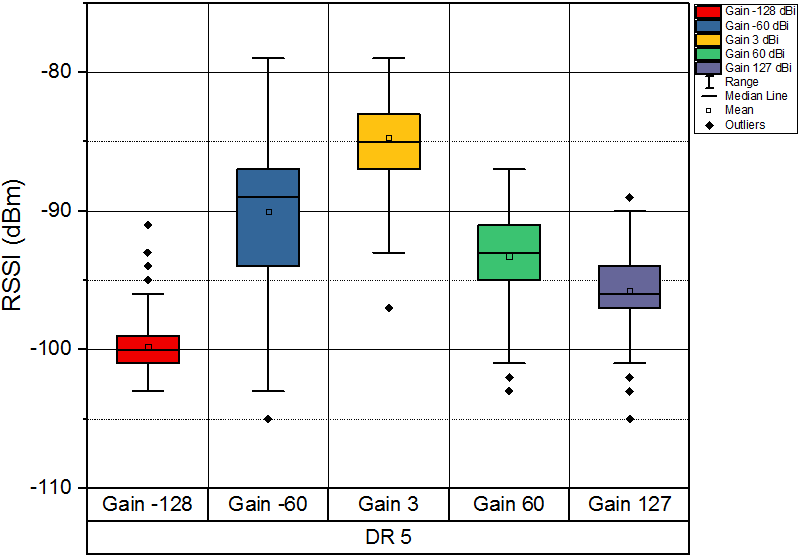}
	\caption[RSSI values by changing the gain]{RSSI values by changing the gain.}
	\label{ch5_f_13}
\end{figure}
\subsubsection{Impact on RSSI by changing the gain}
\label{3}
The end-node (\mdot)~allows a gain of -128 dBi to 127 dBi. For this test, \gls{dr} and transmission power was fixed at \gls{dr} 5 and at 14 dBm respectively. All the readings were taken at the same location. The x-axis represents the value of gain used for testing. Fig.~\ref{ch5_f_13} shows that 3 dBi gain has the strongest signal  because with isotropic antenna and 3 dBi gain the signal is transmitted in all directions. It can be observed that the \gls{rssi} values decrease with either the increased or decrease in gain from 3 dBi. This occurs because, with the increase or decrease of gain from 3 dBi, the signal is transmitted in a particular direction. Thus, the conclusion can be reached that 3 dBi gain should be recommended for use in \lorawan.

\subsubsection{Impact on RSSI by changing the DR in the same location}

\label{1}

This test was performed with different \gls{dr}s. During this test, transmission power was fixed (14 dBm) and also antenna gain was fixed (3 dBi).

\begin{figure}[b]
	\centering
	\includegraphics[width=1.0\linewidth]{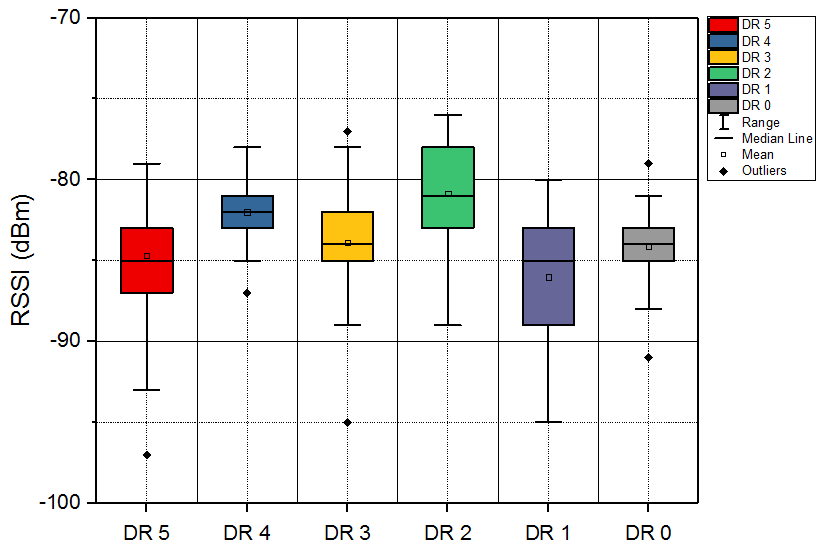}
	\caption[RSSI values measured on the \nth{2} floor at $\approx$ 42 meters distance from the gateway]{RSSI values measured on the \nth{2} floor at $\approx$ 42 meters distance from the gateway.}
	\label{ch5_f_9}
\end{figure}

Fig.~\ref{ch5_f_9} shows the \gls{rssi} values measured for each \gls{dr} at 42 meters distance from the gateway on the same floor. \gls{dr} is represented on the x-axis ranging from \gls{dr} 5 to \gls{dr} 0. From Fig.~\ref{ch5_f_10}, the \gls{rssi} values for all \gls{dr}s are almost same. The values of \gls{rssi} for different \gls{dr}s are in between -80 dBm to -90 dBm. Further, it can be observed that the different \gls{dr} does not have an impact on \gls{rssi}. %while in \cite{petajajarvi2017evaluation, neumann2016indoor}, the authors have also performed the indoor evaluation and conclude the same result, that the different \gls{dr} does not have an impact on \gls{rssi}.

So higher \gls{dr}s would be preferable in order to achieve better performance regarding latency and bit rate without losing the indoor long range coverage for metering applications.

\subsubsection{Impact on RSSI by changing the DR at left and right side from the gateway on the same floor}

\begin{figure}[t]
	\centering
	\includegraphics[width=1.0\linewidth]{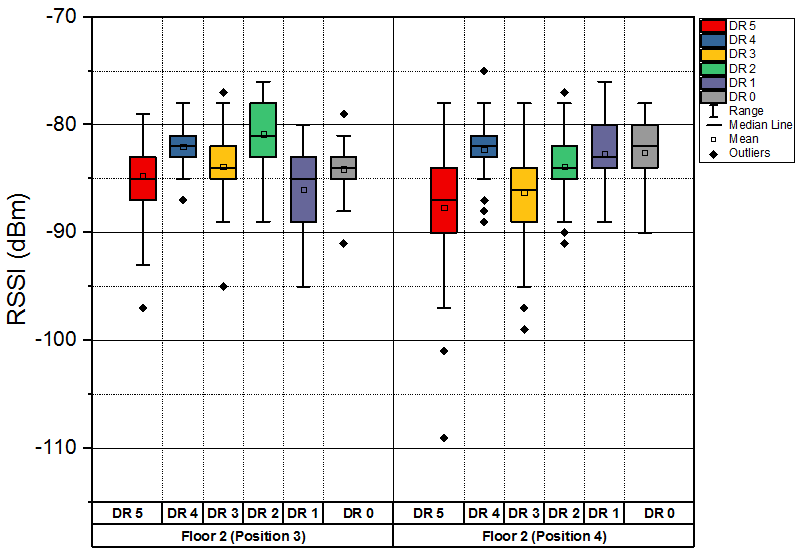}
	\caption[RSSI values measured on \nth{2} floor at left (approx. 42m) and right (approx. 43m) side of the gateway]{RSSI values measured on \nth{2} floor at left ($\approx$ 42 meters) and right ($\approx$ 43 meters) side of the gateway.}
	\label{ch5_f_10}
\end{figure}

Again  for this test, transmission power was fixed (14 dBm) and also antenna gain was fixed (3 dBi).  Fig.~\ref{ch5_f_10} shows the \gls{rssi} values measured for each \gls{dr} at left ($\approx$ 42 meters) and right ($\approx$ 43 meters) side from the gateway on the same floor. In Fig.\ref{ch5_f_10}, the x-axis represents \gls{dr} while the first group shows position 3 and the second group shows position 4. Position 3 and 4 represent left and right of the gateway respectively. From the results shown in the figure, the behavior of the system is the same on the left and right side on the same floor. The \gls{rssi} on both sides for each \gls{dr} is almost equal. This test also concludes that the signal transmitted in all directions has the same strength  with 3 dBi gain. Furthermore, the signal strength is the same in all directions when an isotropic antenna is used with 3 dBi gain.

\subsubsection{Impact on RSSI by changing different floors}

\begin{figure}[t]
	\centering
	\includegraphics[width=1.0\linewidth]{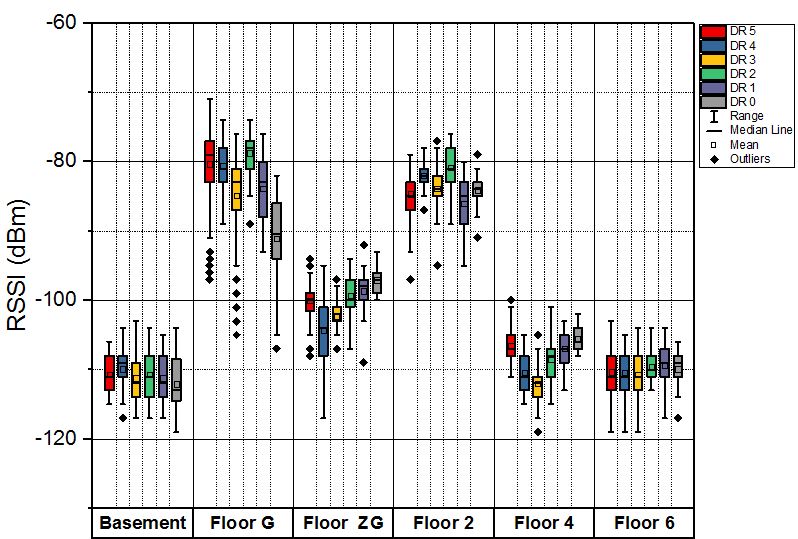}
	\caption[RSSI values on different floors]{RSSI values on different floors.}
	\label{ch5_f_11}
\end{figure}

Generally, the value of \gls{rssi} rapidly decreases in an indoor situation. For this test once again transmission power was fixed at 14 dBm and the gain was fixed at 3 dBi. Table \ref{ch5_t_3} shows the distance between the end-node and the gateway. The positions represent the locations shown in Fig.~\ref{ch5_f_5}. The gateway was located on the \nth{2} floor. Fig.~\ref{ch5_f_11} confirms the results of the first test, i.e. different \gls{dr} does not impact \gls{rssi} values even when the floors are changed. The variance in \gls{dr} values shown in this figure is negligible. 

\begin{table}[h]
	\centering
	\begin{tabular}{ccl}
		\toprule
		\textbf{Floor} & \textbf{Position} & \textbf{\begin{tabular}[c]{@{}l@{}}Distance from\\ the Gateway (m)\end{tabular}} \\ \midrule
		6 & 1 & 51~$\pm$~10 \\ \hline
		4 & 2 & 45~$\pm$~10 \\ \hline
		2 & 3 & 40~$\pm$~3 \\ \hline
		ZG & 5 & 48~$\pm$~10 \\ \hline
		G & 6 & 20~$\pm$~5 \\ \hline
		Basement & 7 & 26~$\pm$~5 \\ \bottomrule
	\end{tabular}
	\caption[End-node distances from the gateway]{End-node distances from the gateway.}
	\label{ch5_t_3}
\end{table}

Because the gateway was on the \nth{2} floor  when the end-node was placed on the same floor the value of \gls{rssi} is higher than the other floors. However, when the end-node was placed on floor 4 and 6 the values of \gls{rssi} were lower because the distance between the gateway and end-node has increased. When the end-node was placed on floor G, the values of \gls{rssi} were higher because this was in the line of sight and there were no obstacles between the end-node and gateway. The \gls{rssi} value decreased for the floor ZG because of an increase in the distance. Lastly, in the basement, the \gls{rssi} value is the lowest not because of the distance (which is almost similar to floor G), but because of the obstacles and concrete structure between the two floors. %Whereas in \cite{neumann2016indoor}, the authors finding are same about the\gls{rssi} on different floor. which is validate our results.

\subsection{Signal-to-Noise Ratio (SNR)}

\gls{snr} defines the ratio between the signal power and the noise power. \gls{snr} is measured in decibels (dB). The quality of an audio signal and transmission channel over the network can be measured by \gls{snr}. It is easier to identify, eliminate and isolate the source of noise if the \gls{snr} value is greater. An original signal cannot be separated from the unwanted noise if the \gls{snr} value is zero. For \gls{snr} there is another abbreviation S/N \cite{Johnson:2006}.

If an \gls{snr} value is greater than 0 dB, it indicates more signal than noise. \gls{snr} is often used metaphorically to point out the ratio of relevant information to incorrect or irrelevant data in an exchange or a conversation \cite{Johnson:2006}.

As the tests performed for \gls{rssi}, the same tests are performed in order to evaluate the behavior of \gls{snr} in the network. For each test, one hour of readings were obtained of the parameters being tested. As expected, the number of packets received by the gateway are increased with the increase in \gls{dr}. In all of the below figures the \gls{snr} is always represented on the y-axis. The complete details of all the tests are given as follows.

\begin{figure}[t]
	\centering
	\includegraphics[width=1.0\linewidth]{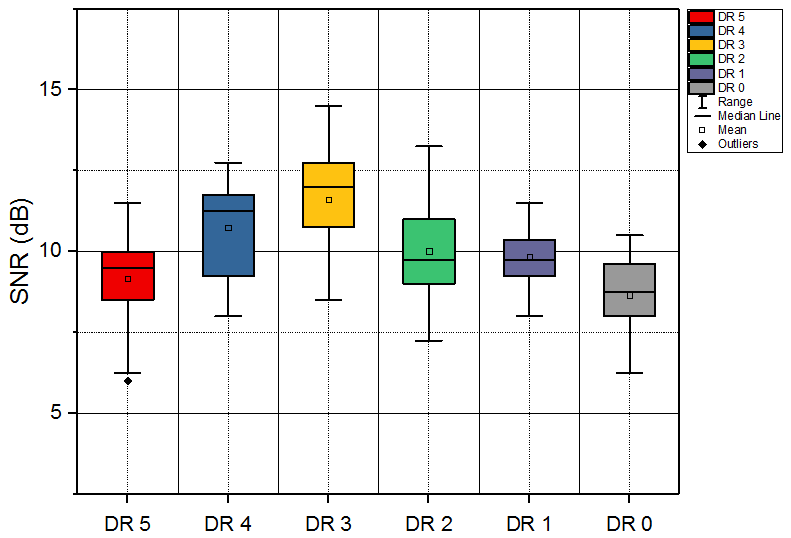}
	\caption[SNR values measured on \nth{2} floor at $\approx$ 42 meters distance from the gateway]{SNR values measured on \nth{2} floor at $\approx$ 42 meters distance from the gateway.}
	\label{ch5_f_14}
\end{figure}
\subsubsection{Impact on SNR by changing the DR in the same location}
This test was performed to assess the impact of \gls{dr} on \gls{snr}. The transmission power was fixed at 14 dBm and antenna gain was fixed at 3 dBi during the test. Location was also not changed for all \gls{dr} during the test and the end-node was on the same floor as the gateway. \gls{dr} is represented on the x-axis and as in the previous section \ref{1}, the colored part represents 25\% to 75\% of the values with the whiskers showing the complete range of values. The small clear square represents the mean, while the line inside each of the box represents the median. The outliers are represented by the diamond shape outside the whiskers. 

Fig.~\ref{ch5_f_14} shows the value of \gls{snr} with  the changes in \gls{dr}. It can be observed that the changes in \gls{dr} bring minor changes in \gls{snr} specifically the \gls{snr} is higher with \gls{dr} 3. Other than this there are no big changes  occurred with changing the \gls{dr}. The range of \gls{snr} falls between 7.5 to 12.5 dB. It can be concluded from this test that changes in \gls{dr} have an impact on \gls{snr}.

\subsubsection{Impact on SNR by changing the DR at left and right side from the gateway on the same floor}

\begin{figure}[b]
	\centering
	\includegraphics[width=1.0\linewidth]{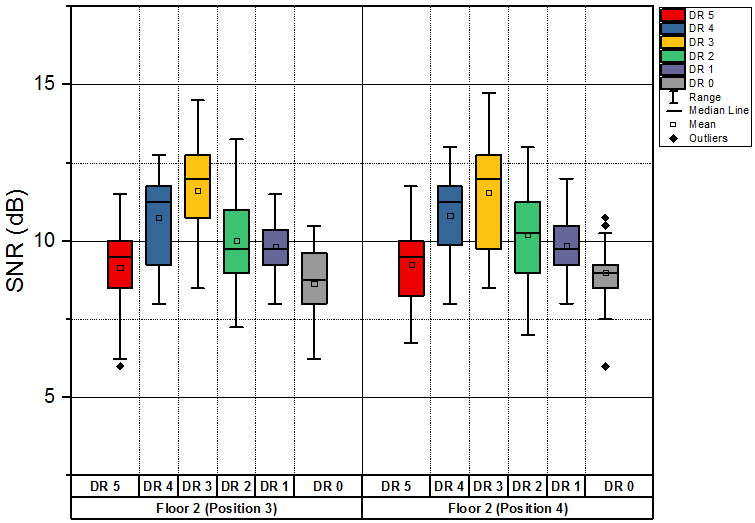}
	\caption[SNR values measured on the \nth{2} floor at left (approx. 42 meters) and right (45 meters) side of the gateway]{SNR values measured on the \nth{2} floor at left (approx. 43 meters) and right (45 meters) side of the gateway.}
	\label{ch5_f_15}
\end{figure}

For this test, once again transmission power was fixed at 14 dBm and the gain was fixed at 3 dBi. %As with the previous tests
The distance of the gateway from the end-node was $\approx$ 42 meters to the left and $\approx$ 43 meters to the right. The x-axis represents the \gls{dr} and the first group shows position 3 and the second group shows position 4. Position 3 and 4 represent the left and right of the gateway respectively. Fig.~\ref{ch5_f_15} shows that the performance of the network is same on both sides of the gateway. This test further concludes that the \gls{snr} remains same in all directions with 3 dBi gain.

%\pagebreak
\subsubsection{Impact on SNR by changing different floors}

The distance between the gateway and the end-node along with floor location is shown in Table \ref{ch5_t_3}.  For this test once again transmission power was fixed at 14 dBm and the gain was fixed at 3 dBi.

Because the gateway was located on the \nth{2} floor the signal quality is best on the same floor. However, \nth{4} floor shows that \gls{snr} values have decreased due to increase in distance and obstacles. Since the end-node located on the G floor was in the line of sight, the \gls{snr} value here is also very well similar to the end-node located on \nth{2} floor. Floor ZG has an increased distance which is almost comparable to \nth{4} and \nth{6} floor but the \gls{snr} value is almost as high as floor G. Although, the distance has increased there are less physical obstacles between the gateway and the end-node. The \gls{snr} value is lowest at basement and \nth{6} floor due to the fact that there is an increased distance on the \nth{6} floor but in the basement, there is a lot of hindrance because of the concrete structure. It can be concluded from these results that the networks in the basement and for long distances should be configured with special attention.
\begin{figure}[t]
	\centering
	\includegraphics[width=1.0\linewidth]{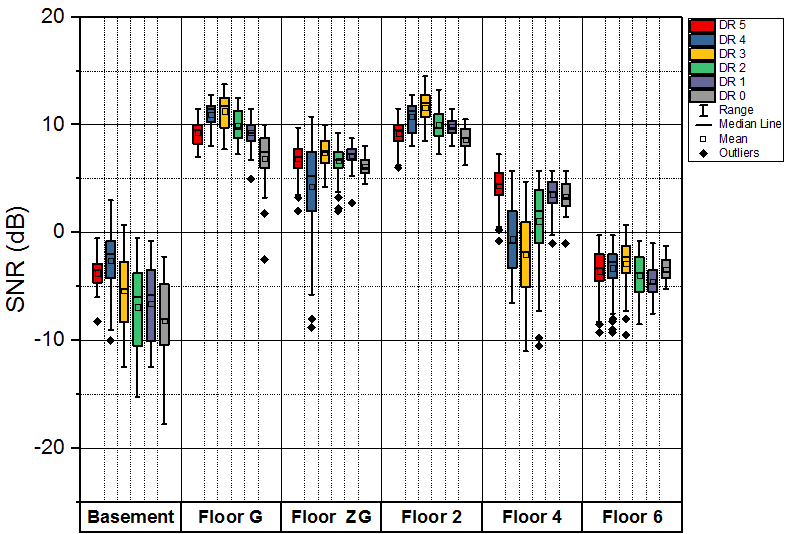}
	\caption[SNR values on different floors]{SNR values on different floors.}
	\label{ch5_f_16}
\end{figure}
\begin{figure}[b]
	\centering
	\includegraphics[width=1.0\linewidth]{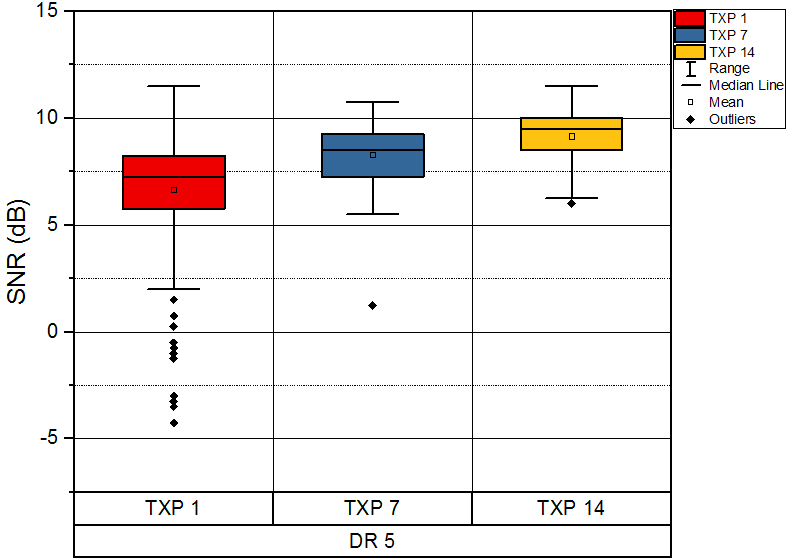}
	\caption[SNR values by changing the transmission power]{SNR values by changing the transmission power.}
	\label{ch5_f_17}
\end{figure}
\subsubsection{Impact on SNR by changing the transmission power}

As previously explained in section \ref{3} the end-node allows a transmission power of 1 dBm to 30 dBm. During  the test, the range from 1 dBm to 14 dBm of power transmission was used. The test was performed with a fixed \gls{dr} of \gls{dr} 5 and gain of 3 dBi. Fig.~\ref{ch5_f_17} shows a clear trend of increase in signal strength with the increase in transmission power. Because of this increase in signal strength, the original signal becomes stronger in comparison to the noise and thus produces a better \gls{snr} value.

\subsubsection{Impact on SNR by changing the gain}
\begin{figure}[t]
	\centering
	\includegraphics[width=1.0\linewidth]{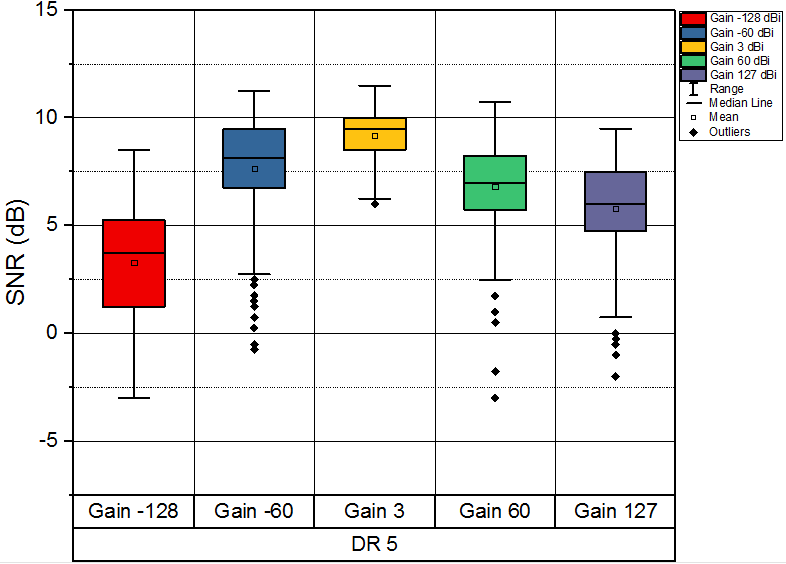}
	\caption[SNR values by changing the gain]{SNR values by changing the gain.}
	\label{ch5_f_18}
\end{figure}

As explained previously in section \ref{3} the end-node allows a gain of -128 dBi to 127 dBi. For this test, \gls{dr} and transmission power was fixed at 14 dBm respectively. All the readings were taken at the same location. As always x-axis represents the value of gain. Fig.~\ref{ch5_f_18} shows that with 3 dBi gain the best \gls{snr} values is achieved. This is due to the fact that 3 dBi gain transmits equal strength signal in all directions. However, with the increase or decrease in gain, the signal is transmitted in a specific direction. Because of this, the quality of signal weakens and the noise level increases, which in turn produces a lower \gls{snr} value. 

\subsection{Impact of different \lorawan~classes on Acknowledgment (ACK)}

This test was performed to check the behavior of different \lorawan~classes in terms of downlink communication. As described in section \ref{classes}, a class A type end-node opens two short windows for downlink communication and class C type end-node continuously listens for downlink communication until next uplink packet. For this test, all the parameters were same for \gls{dr} 5 and \gls{dr} 4 only class type was changed.

\begin{table}[h]
	\centering
	\begin{tabular}{ccc}
		\toprule
		\textbf{DR} & \textbf{\begin{tabular}[c]{@{}c@{}}End-node Class\end{tabular}} & \textbf{\begin{tabular}[c]{@{}c@{}}Acknowledgment (ACK) missed\end{tabular}} \\ \midrule
		5 & A & 50 \\ \hline
		5 & C & 23 \\ \hline
		4 & A & 42 \\ \hline
		4 & C & 21 \\
		\bottomrule
	\end{tabular}
	\caption[Acknowledgment (ACK) received with different \lorawan~classes]{Acknowledgment (ACK) received with different \lorawan~classes.}
	\label{ch5_t_4}
\end{table}

As a result of Table \ref{ch5_t_4}, it can be seen that for class C end-node received 50\% more \gls{ack} than class A. So class C end-node would be preferable in low latency application.

\subsection{Packet Error Rate (PER)}

The \gls{per} refers to the packets, which are not received by the gateway. The results are summed up in Table \ref{ch5_t_2}. In Table \ref{ch5_t_2}, we list the distance between the gateway and each end-node location as well as position number that has been given to that location. The location of positions is shown in Fig.~\ref{ch5_f_5}. Lastly, the table shows the \gls{per} of each position for different \gls{dr}s. During this test, transmission power was fixed (14 dBm) and also antenna gain was fixed (3 dBi).

\begin{table*}[h]
	\centering
	\begin{tabular}{lllllllll} 
		\toprule
		\multicolumn{7}{r}{\textbf{Packet Error Rate (\%)}}\\ 
		%\midrule
		\textbf{Floor} & \textbf{\begin{tabular}[c]{@{}c@{}}Position\end{tabular}} & \textbf{\begin{tabular}[c]{@{}c@{}}Distance from the Gateway(m)\end{tabular}} & \textbf{DR 5} & \textbf{DR 4} & \textbf{DR 3} & \textbf{DR 2} & \textbf{DR 1} & \textbf{DR 0} \\ 		\midrule
		6 & 1 & 51~$\pm$~10 & 65 & 2 & 0 & 0 & 0 & 0 \\ \hline
		4 & 2 & 45~$\pm$~10 & 2 & 1 & 8 & 1 & 0 & 0 \\ \hline
		2 & 3 & 40~$\pm$~3 & 0 & 4 & 0 & 2 & 0 & 0 \\ \hline
		2 & 4 & 42~$\pm$~3 & 0 & 0 & 0 & 0 & 0 & 0 \\ \hline
		ZG & 5 & 48~$\pm$~10 & 2 & 4 & 1 & 0 & 0 & 0 \\ \hline
		G & 6 & 20~$\pm$~5 & 0 & 0 & 0 & 1 & 0 & 0 \\ \hline
		B & 7 & 26~$\pm$~5 & 90 & 28 & 10 & 5 & 15 & 11 \\ \hline
		B & 8 & 60~$\pm$~10 & 100 & 100 & 100 & 100 & 100 & 100 \\ \hline
		B & 9 & 55~$\pm$~10 & 100 & 100 & 100 & 100 & 100 & 100 \\ \bottomrule
	\end{tabular}
	\caption[Packet Error Rate (PER) of different data rate from different positions]{Packet Error Rate (PER) of different data rate from different positions.}
	\label{ch5_t_2}
\end{table*}

On average \gls{per} is 0 to 4\% except for the basement. On the \nth{6} floor and in the basement, 65\% and 95\% of packets are missed respectively for \gls{dr} 5 due to long distance and obstacles. Thus for long distances, higher \gls{sf} is preferable to minimize the \gls{per}. In the basement, the last two positions did not receive any packets due to the concrete structure and the long distance. From these results, it can be concluded that with the decrease of \gls{dr} the \gls{per} will also decrease. That is why the selection of \gls{dr} is important in order to maximize the performance of the network.

\subsection{\lora~Network Capacity}

Practically it is impossible to find out network capacity with a single end-node and gateway. This is the limitation of our study, we have calculated \lora~network capacity theoretically.

In a \lora~network, the class A devices open two short windows for downlink communication at 1 sec and 2 sec after uplink communication. In order to calculate the \lora~network capacity, let's assume it will take 2 sec to send one packet and get a response from the network server. Therefore, in 1 hour, there are ($\frac{60 \times 60}{2}$) or 1800 packet opportunities. This puts the daily capacity of the network traffic at 43200 packets. The Nodes at the edge of the network needs more time to transmit the packets, so double time is required for counting the edge nodes %because of extended \gls{toa} 
\cite{cattani2017experimental}. Now if the gateway has one channel then the total number of nodes can  be connected per day to the gateway.

\begin{align}
Number~of~nodes= NR = \frac{43200}{R + ER \times 2}  
\end{align}
\begin{itemize}
	\item R - Packets per day requiring a response.
	\item ER - Edge packets per day requiring a response. 
\end{itemize}
However, in our case, we are using the MultiConnect Conduit gateway. This gateway can receive 8 packets simultaneously if these packets are sent using  different frequencies and \gls{sf}s. So, for the Conduit gateway, the total number of nodes are
\begin{align}
	\label{e_2}
	NR= \frac{345600}{R + ER \times 2}
\end{align}
These are just base numbers, variability still exists in the \gls{dr} and the size of data to be sent. Nodes at a further distance will need more \gls{toa} and transmission power to send the same number of bytes, due to the long distance. %So the closer nodes are to the gateway the higher the capacity. %In order to calculate the network usage of Conduit gateway capacity we can use the following equation.
%\begin{align}
%\label{e_3}
%Network~Usage = NR ( R + ER * 2) / 345600
%\end{align}

From equation \ref{e_2} if 14400 nodes require \gls{ack} and each node sends 24 packets per day than 100\% of the network will be used. If we send 10 packets per node per hour. Usually, for 1 day, we can send 240 packets per node per day. So maximum 1440 nodes can be connected to our Conduit gateway. 

From the above results, we can conclude that the network capacity of \lora~depends upon several parameters like the number of channels of the gateway, if \gls{ack} is needed or not, and the number of packets required per day.

\section{Results of the Outdoor evaluation of the \lora~protocol}
\label{outdoor}
We conducted outdoor testing at the University of Applied Sciences\footnote{https://www.fh-ooe.at/en/hagenberg-campus/} Hagenberg Campus, Austria. We placed the gateway in a laboratory (PL3) located at the top floor in the FH 2 building. The location of the gateway for the outdoor scenario is shown in Table \ref{ch5_t_5}. The gateway was placed in the lab, but its antenna was mounted outside the window with the help of an extension cable. We selected this location because the university is located on higher ground as compared to the surrounding area. Other advantages of this location are that on one side of the university there are many buildings and obstacles, and on the other side, there is a straight line of sight.

\begin{table}[h]
	\centering
	\begin{tabular}{ll}
		\toprule
		\textbf{Location} & \textbf{FH-OOE} \\ \midrule
		Latitude & 48°22'5.00"N \\ \hline
		Longitude & 14°30'46.70"E \\ \hline
		Sea Level & 458.12 \\ 
		\bottomrule
	\end{tabular}
	\caption[Location of the gateway for outdoor testing]{Location of the gateway for outdoor testing\protect\footnotemark.}
	\label{ch5_t_5}
\end{table}

\footnotetext{\url{www.google.at/maps/}\label{fn2}}

For outdoor testing, the end-node was placed at three different locations which were provided generously by some colleagues at their homes. We used the same application for the end-node to generate the dummy data, which is described in section \ref{applicaiton}. 
Different tests have been performed at different times and days. The outdoor tests are as follows.

\subsection{\lorawan~Coverage}

\begin{figure}[h]
	\centering
	\includegraphics[width=1.0\linewidth]{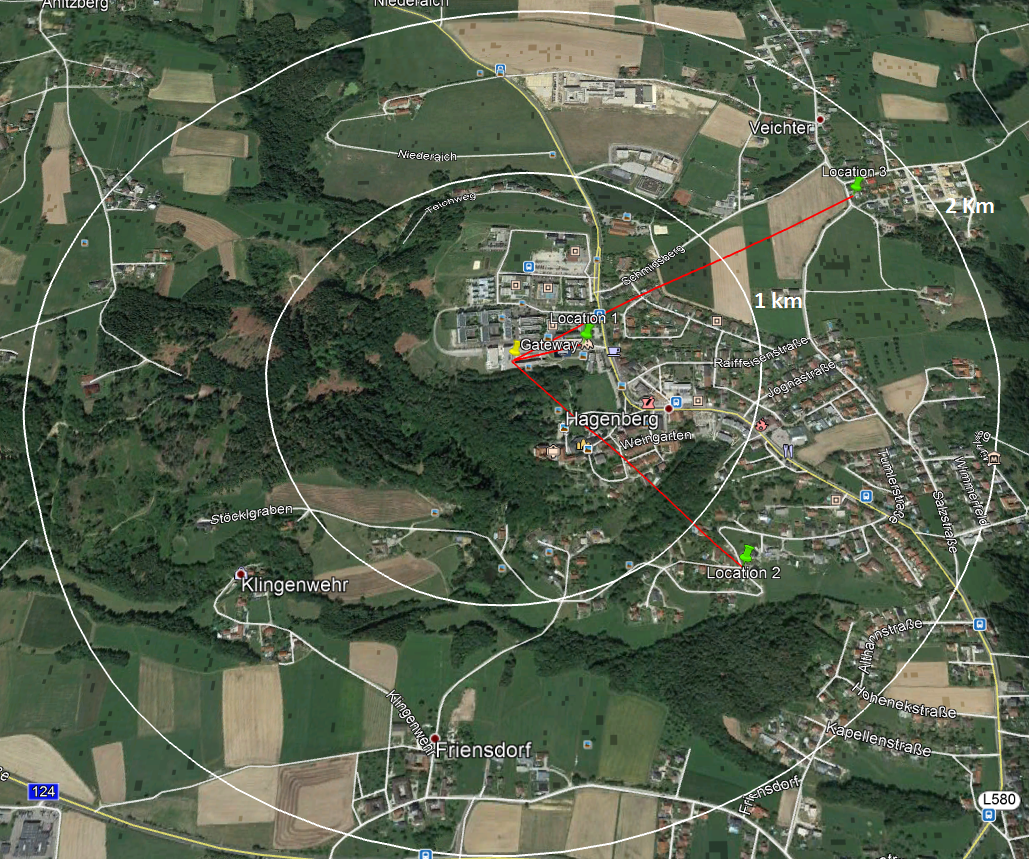}
	\caption[Outdoor locations of the gateway and end-node]{Outdoor locations of the gateway and end-node\protect\footnotemark.}
	\label{ch5_f_19}
\end{figure}

One of the most important aspects of wireless technologies is to determine their coverage. % \lorawan~is the coverage range. 
This is very crucial for the cost estimation of the network. For this purpose, we performed a range test of \lorawan~in Hagenberg, Austria. The aim of this test was to check the range of \lorawan~with a single gateway for the smart metering application. We selected 3 different locations around the gateway with three different distances as shown in Fig~\ref{ch5_f_19}. The coordinates and distance of each location from the gateway is shown in Table \ref{ch5_t_6}.

\begin{table}[h]
	\centering
	\begin{tabular}{llll}
		\toprule
		\textbf{Location} & \textbf{Co-ordinates} & \textbf{\begin{tabular}[l]{@{}l@{}}Dist. from\\the Gateway\\(m)\end{tabular}} & \textbf{\begin{tabular}[l]{@{}l@{}}Height from\\Sea Level\\(m)\end{tabular}} \\ \midrule
		1 & 48°22'6.24"N  14°30'53.81"E & 151.14 & 463.9 \\ \hline
		2 & 48°21'50.60"N 14°31'8.80"E & 635.8 & 402.64 \\ \hline
		3 & 48°22'18.60"N  14°31'22.30"E & 845 & 472.44 \\ 
		\bottomrule	
	\end{tabular}
	\caption[End-node locations and distances from the gateway]{End-node locations and distances from the gateway.}%\footref{fn2}.}
	\label{ch5_t_6}
\end{table}
%\footnotetext{\url{www.google.at/earth/}\label{fn2}}}
\footnotetext{\url{www.google.at/earth/}}
The transmission power was fixed at 14 dBm, while antenna gain was fixed at 3 dBi. The results of our experiment show that the maximum coverage of a single gateway is about 1.70 km in radius. When the end-node was placed on location 1 (closest location), the gateway received all the packets with \gls{dr} 0--5. However, when the end-node was placed on maximum distance (location 3), the gateway received fewer packets with \gls{dr} 2 and 3 but with \gls{dr} 0, the gateway received almost all the packets. The distance of location 2 was less than location 3, but the gateway only received the packets with \gls{dr} 0, due to the obstacles, which included buildings and lots of trees.

From the above results, we can conclude that network coverage of \lorawan~depends upon several parameters like transmission power, antenna direction, height and obstacles etc.

\subsection{863--870 MHz Band Usage}

For this test, one hour of reading was acquired for each \gls{dr}. Altogether, we took 6 hours of recording the data with \gls{dr} 0-5. During 6 hours, a total number of 773 packets were received by the gateway.  

\begin{figure}[h]
	\centering
	\includegraphics[width=1.0\linewidth]{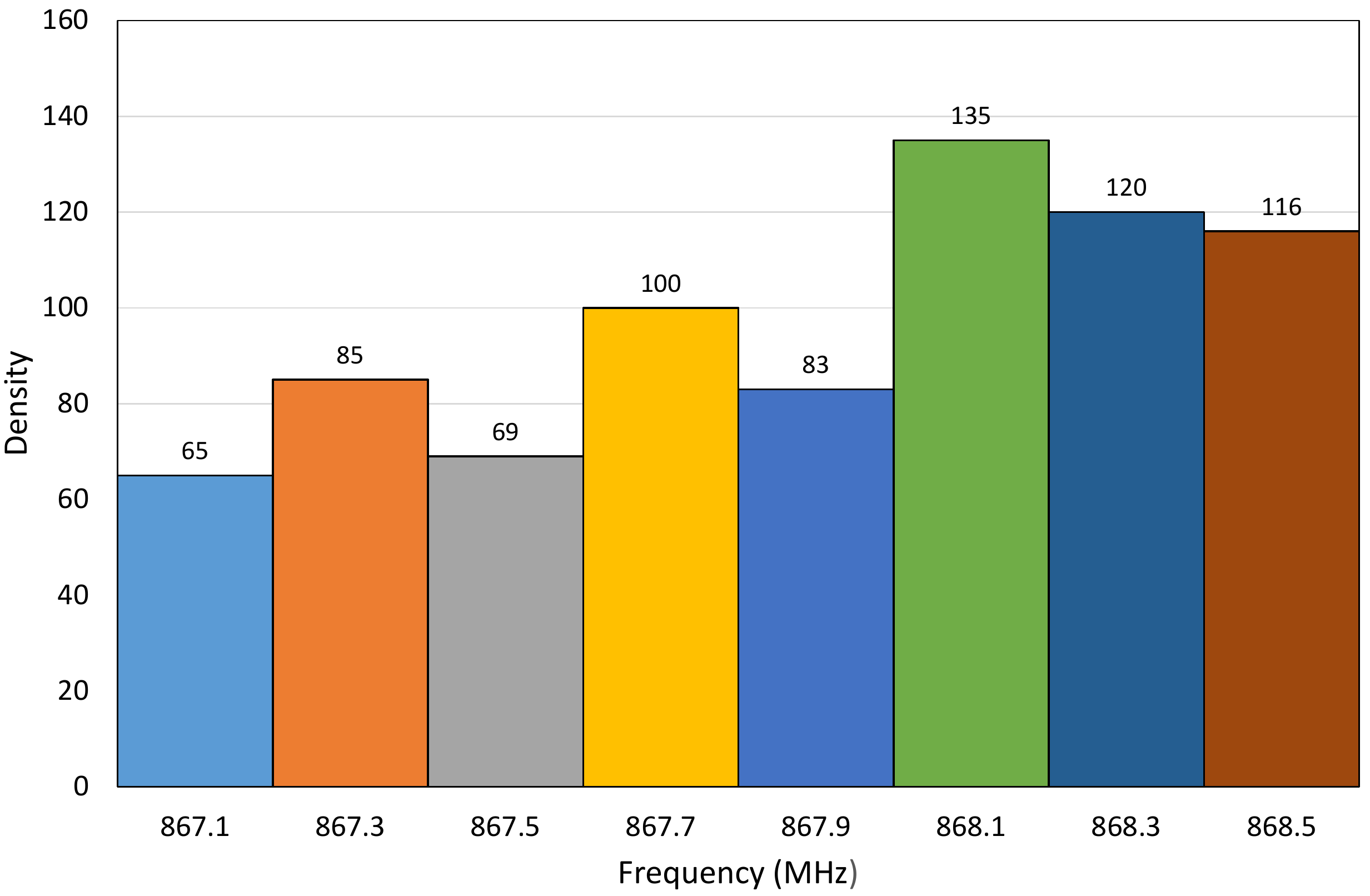}
	\caption[Histogram of 863-870 MHz band usage (total packets 773)]{Histogram of 863-870 MHz band usage (total packets 773).}
	\label{ch5_f_20}
\end{figure}

In the EU region, \lorawan~operates in 863-870 MHz band. In our case, we are using the MultiConnect Conduit gateway. This gateway can receive 8 packets simultaneously if they are using  different frequencies and \gls{dr}s. Fig.~\ref{ch5_f_20} shows the frequencies, which were encountered in the test. The 3 mandatory frequencies/channels 868.1 MHz, 868.3 and 868.5 were used the most. The other 5 frequencies are supporting frequencies in order to increase the capacity of the network. 

\section{Conclusion}
\label{conclusion}

In this paper, a \lorawan~is implemented with Multitech devices (end-node and gateway) in order to evaluate the performance of the network for indoor and outdoor realistic scenarios, under the European regulations. From our experiences, \lorawan~is easy to setup and configure for real-time smart applications like smart meter. Our results show that a single gateway with eight sub-band channels
is able to receive different uplink and downlink packets simultaneously with different DRs and frequencies within a radius of 1.70 km. As smart meter sends 96 bytes of data
during different times in a day, thus we can easily receive data from thousands of smart meters, with a single gateway. In other words, a gateway can handle thousands of smart meters at a very low cost. While deploying the networks in the urban area
one should pay special attention to the DR, position, and the height of the antenna in order to increase the performance and range. It can also be concluded from our research, different DRs do not have an impact on the
signal strength. The quality of the signal depends on the distance from the gateway, the
number of obstacles, transmission power, antenna type, antenna height, and antenna gain.
Furthermore, for scenarios where a lot of concrete buildings and obstacles are present, based on our results we recommend to use an isotropic antenna with the maximum allowed transmission power value. If the end-node is very close to the gateway, then
low transmission power is preferable in order to achieve better signal strength.

\section*{Acknowledgment}
The authors would like to thank building management of the Johannes Kepler University, and the University of Applied Sciences Upper Austria for giving us access to the buildings and allowing us to conduct the studies.  

\ifCLASSOPTIONcaptionsoff
  \newpage
\fi

%\begin{thebibliography}
\bibliographystyle{IEEEtran}
\balance
\bibliography{references}

\end{document}